\documentclass[twocolumn,floats,floatfix,prl,nofootinbib]{revtex4}
\usepackage{bm}
\usepackage{graphicx}
\usepackage{amsmath}
\usepackage{amsfonts}
\usepackage{mathrsfs}

\usepackage[utf8]{inputenc}
\usepackage{tabularx}

\usepackage{xcolor}
\usepackage[normalem]{ulem}
\newcommand\strike{\bgroup\markoverwith{\textcolor{red}{\rule[.5ex]{2pt}{1pt}}}\ULon}

\def\ph{+\frac{1}{2}}
\def\pmh{\pm \frac{1}{2}}
\def\mh{-\frac{1}{2}}
\newcommand{\hslashslash}{%
  \raisebox{.85ex}{%
    \scalebox{.7}{%
      \rotatebox[origin=c]{18}{$-$}%
    }%
  }%
}
\newcommand{\bbar}{%
  {%
   \vphantom{b}%
   \kern0.09em
   \ooalign{\kern-0.16em\smash{\hslashslash}\hidewidth\cr$b$\cr}%
   \kern0.05em
  }%
}

\begin{document}

\title{Composite 
Dislocations in Smectic Liquid Crystals}

\author{Hillel Aharoni}
\affiliation{Department of Physics and Astronomy, University of Pennsylvania, 209 South 33rd Street, Philadelphia, Pennsylvania 19104, USA}
\author{Thomas Machon}
\affiliation{Department of Physics and Astronomy, University of Pennsylvania, 209 South 33rd Street, Philadelphia, Pennsylvania 19104, USA}
\author{Randall D. Kamien}\email{kamien@upenn.edu}
\affiliation{Department of Physics and Astronomy, University of Pennsylvania, 209 South 33rd Street, Philadelphia, Pennsylvania 19104, USA}

\date{\today}

\begin{abstract}

Smectic liquid crystals are charcterized by layers that have a preferred uniform spacing and vanishing curvature in their ground state. Dislocations in the smectics play an important role in phase nucleation, layer reorientation, and dynamics.
Typically modeled as possessing one line singularity, the layer structure of a dislocation
leads to a diverging compression strain as one approaches the defect center, suggesting a large, elastically determined melted core. However, it has been observed that for large charge dislocations, the defect breaks up into two disclinations [C. E. Williams, Philos. Mag. {\bf 32}, 313 (1975)].  Here we investigate the topology of the composite core.  Because the smectic cannot twist, transformations between different disclination geometries are highly constrained.
We demonstrate the geometric route between them and show that despite enjoying precisely the topological rules of the three-dimensional nematic, the additional structure of line disclinations in three-dimensional smectics localizes transitions to higher-order point singularities.
\end{abstract}

\maketitle
Dislocations are, by their nature, not only topological but {\sl geometrical}: by definition, they only occur in systems with broken translational order and therefore they must induce strain in the crystal or liquid crystal that host them~\cite{klemanbook,mermin,ack}.  These strains can grow quite large and often require a cutoff at the core to keep the energy finite.  In exchange, the core melts into a higher-symmetry phase bringing with it the higher energy of the uncondensed condensate.  Screw dislocations are especially troublesome because of a geometric consequence of their topology. Namely, the helicoidal layer structure that makes up the screw disclocation is not measured at its core~\cite{poenaru}, that is, all the layers come together on the centerline. It follows that the compression energy must diverge there~\cite{msk,kl}. The symmetry of the smectic phase allows the core regions of a dislocation to be replaced by disclination pairs, for both edge and screw~\cite{meyer10,achard05,klbook,williams75} dislocations.
Recall that line defects in nematics are characterized only by a $\mathbb{Z}_2=\pi_1(\mathbb{R}P^2)$ charge; however, when the director lies in the plane perpendicular to the defect line we can assign a geometric charge
In this paper we discuss this phenomenon, and elucidate the topology that allows an edge dislocation can become a screw dislocation through the conversion of a disclination with the a $\ph$ geometry transforms into a $\mh$ geometry.

Before considering composite cores, we first compare the energetic situation in smectics with the theory of superconductors.  Though the harmonic theory of smectics matches the London theory of superconductors~\cite{London,LL} and the Landau theories are strikingly similar~\cite{dgsmectic}, the nonlinear elasticity of the smectic, required by rotational invariance~\cite{halseynelson} captures both the geometry and the diverging energy density of a screw defect.  We locate the smectic layers as level sets of a three-dimensional phase field $\phi({\bf x})$.  This is locally the phase of a complex scalar order parameter, $\psi=\langle\exp\{iq\phi\}\rangle$ where $2\pi/q$ is the equilibrium smectic spacing.  The elastic free energy is
\begin{equation}\label{eq:elas_en}
F=\frac{B}{8}\int d^3\!x\,\left\{\left[\left(\nabla\phi\right)^2-1\right]^2 + 16\lambda^2 H^2\right\},
\end{equation}
where $B$ is the bulk modulus, $\lambda$ is the bend penetration depth, $H=\frac{1}{2}\nabla\cdot(\nabla\phi/\vert\nabla\phi\vert)$ is the mean curvature of the level sets, and we have set $q=1$ for simplicity. 

\begin{figure}[t]
\centerline{\includegraphics[width=0.49\textwidth]{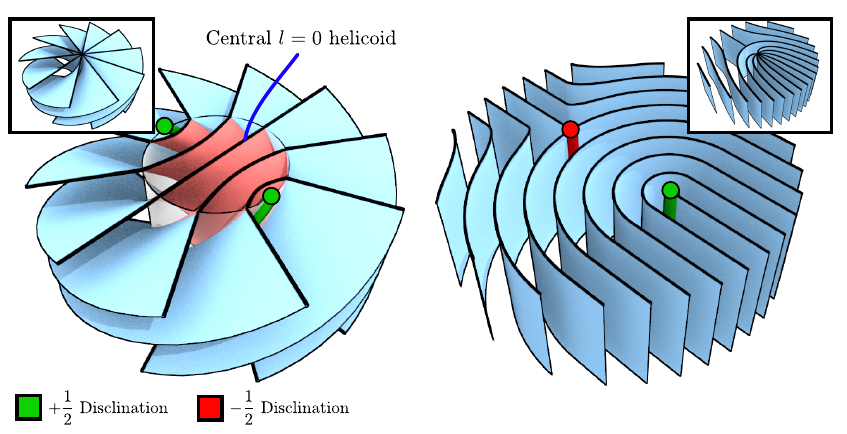}}
\caption{\label{fig:composite}
	{(Left)} A composite smectic screw dislocation. The inner (red) core has zero compression, and costs only a finite bending energy. The outer (cyan) shell has zero mean curvature, and costs only a finite compression energy. The structure, with Burgers scalar $b=10$ (in units of the layer spacing), is completed by the two helical (green) $+\frac{1}{2}$ line disclinations at $b$ half-layers (density minima) away from each other. The radius of the core is $\protect\bbar=b/(2\pi)$. 
	{(Right) A composite edge dislocation. The structure is composed of a $\ph$/$\mh$ disclination pair, the number of half-layers between them again sets $b=10$. Here the mean curvature does not vanish outside the core, nonetheless disclinations are still found at curvature singularities.
	Insets: The standard dislocation structures of equivalent Burgers scalars, with diverging core energy.}
}
\end{figure}

There is a compact three-dimensional set of ground states, $\overline\phi = {\bf n}\!\cdot\!{\bf x} + \phi_0$ parameterized by a unit vector ${\bf n}\in S^2$ and a scalar global phase $\phi_0\in S^1$. Note that the ground state manifold is further reduced by the discrete nematic symmetry ${\bf n}\rightarrow -{\bf n}$, to form a twisted circle bundle over $\mathbb{R}P^2$. Typically, one expands around the ground state $\phi=\overline\phi -u$ in the Eulerian co\"ordinate $u$ to find the harmonic theory~\cite{LL}
\begin{equation}
F_{\rm harm}=\frac{B}{2}\int d^3\!x\,\left\{\left({\bf n}_0\!\cdot\!\nabla u \right)^2 + \lambda^2\left(\nabla^2 u\right)^2\right\}.
\end{equation}
It is remarkable that a screw dislocation with Burgers scalar~\cite{foot} $b$, $\phi_{\rm screw} = z - \frac{b}{2\pi}\tan^{-1}(y/x)$, an extremal of both the full and harmonic free energy functionals, has vanishing energy density in the harmonic theory but a diverging energy density in the rotationally-invariant theory scaling as $Bb^4/r^4$ with $r^2=x^2+y^2$. The linear elasticity theory is a poor starting point for an energetic description. For edge dislocations, the situation is almost opposite: the linear and nonlinear theories give different layer structures but the same energy~\cite{BPS}. 

However, symmetry offers a way out: the $\nabla\phi \to -\nabla\phi$ symmetry of the smectic phase means that $\phi$ lives in the quotient space
$S^1/\mathbb{Z}_2\cong\left\{\mathbb{R}: \phi\sim\phi + 2\pi, \phi \sim -\phi\right\}$ \cite{ack}.
It is important to note that this space is {\sl not} $\mathbb{R}P^1$ where $\phi$ and $\phi+\pi$ are identified and which is not simply connected -- the action of the smectic symmetry on $\phi$ is not free, it has two fixed points: $0$ and $\pi$ -- the layers and the ``half layers''. The level sets that correspond to layers or half-layers must be at these fixed points for a single valued density field $\rho\propto\cos\phi$, so that disclinations must lie on density minima or maxima. This condition breaks the continuous symmetry, $\phi\rightarrow\phi + {\rm constant}$, and generates a Peierls-Nabarro barrier to dislocation glide \cite{slse}. This extra structure allows dislocation cores to split into disclinations: an initial phase singularity of $2\pi$ is equivalent to a phase change from $0$ to $\pi$ followed by the reverse change from $\pi$ to $0$ since the sign of $\nabla\phi$ changes at the fixed points. This process removes the phase singularity but preserves the phase winding that signifies the dislocation.

This fact allows the system to replace the high energy cores of standard dislocations. Such composite dislocations can be described in terms of an almost-equally-spaced structure as described by Kleman and co-workers~\cite{meyer10,achard05}. The topology of a screw dislocation requires the solution $\phi_{\rm screw}$ at large distances, a solution with vanishing mean curvature $H$, however one can replace the divergent-compression core with equally spaced layers (Fig.\ \ref{fig:composite}). Such a core is built with layers specified as the normal evolution of a central helicoid (discussed in detail below). These layers are equally spaced, but are not minimal surfaces, and there are two curvature singularities created at a radius equal to the reduced Burgers scalar, $\bbar = b/2\pi$, which form a double helix. These singularities are the location of the disclinations, and indeed these double helices were observed in screw dislocations with large (giant) Burgers scalar~\cite{williams75,meyer10}.
Outside the core, we attach helicoids to the helices which bound the developed layers. Each of these helices serves as seed for the helicoidal layer outside the core.  We will return to details of this construction in the following.  The Burgers scalar of such a split dislocation is again determined by the number of layers between the disclinations (Fig.~\ref{fig:composite}).

\begin{figure}
	\centerline{\includegraphics[width=0.4\textwidth]{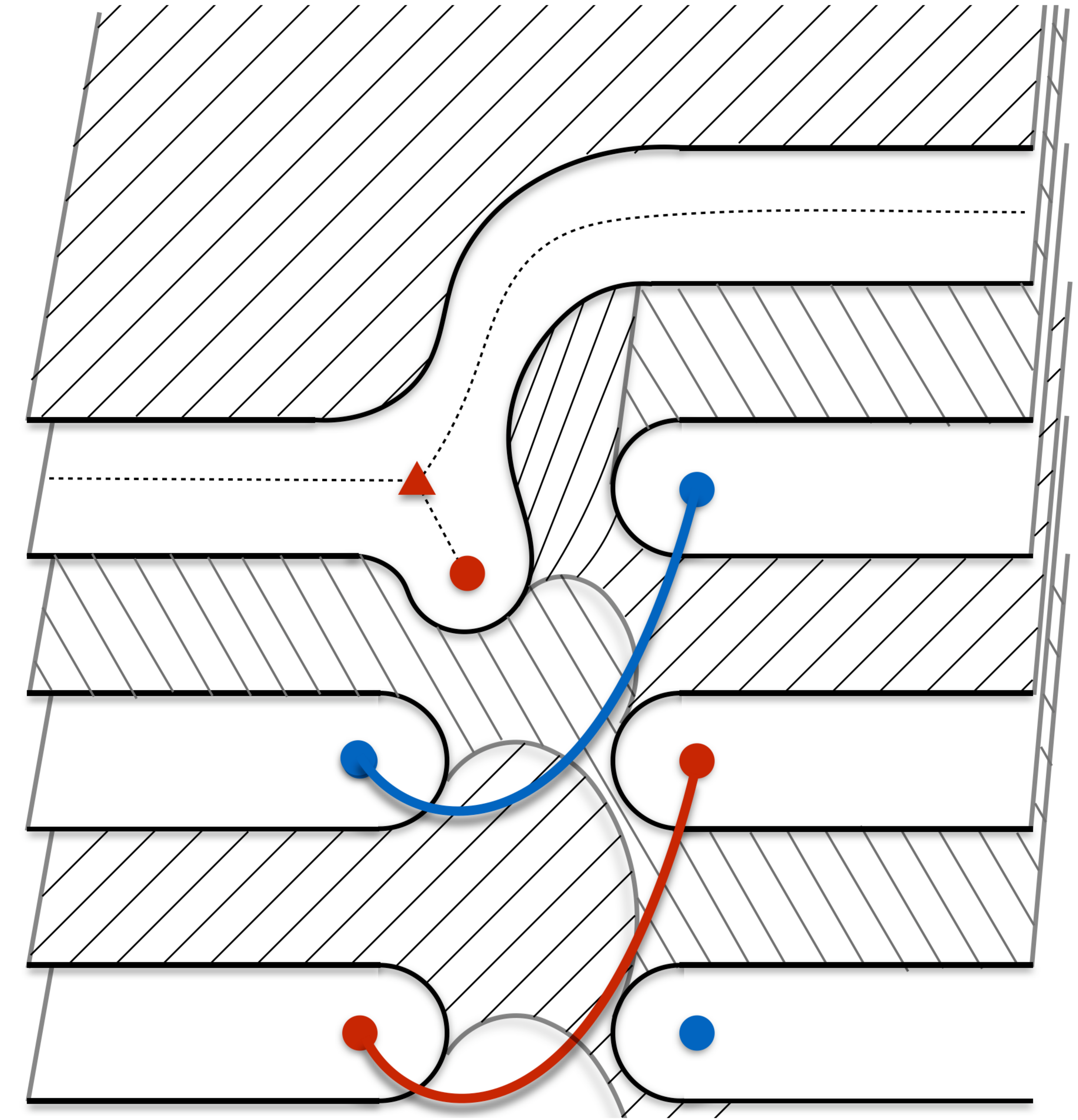}}
	\caption{\label{fig:screw2edge} A composite screw dislocation becoming a composite edge dislocation (with Burgers scalar 4 times the layer spacing).  Note that the a {\sl pincement} is made on the top layer,  a $\mh$/$\ph$ disclination pair with no net dislocation charge~\cite{klemanbook,ack} (red triangle and red circle, respectively).  The ``half layer'' is indicated as a dotted line to show the the pinch. The red and blue circles are the $\ph$ disclinations.
	}
\end{figure}

\begin{figure}
	\centerline{\includegraphics[width=0.49\textwidth]{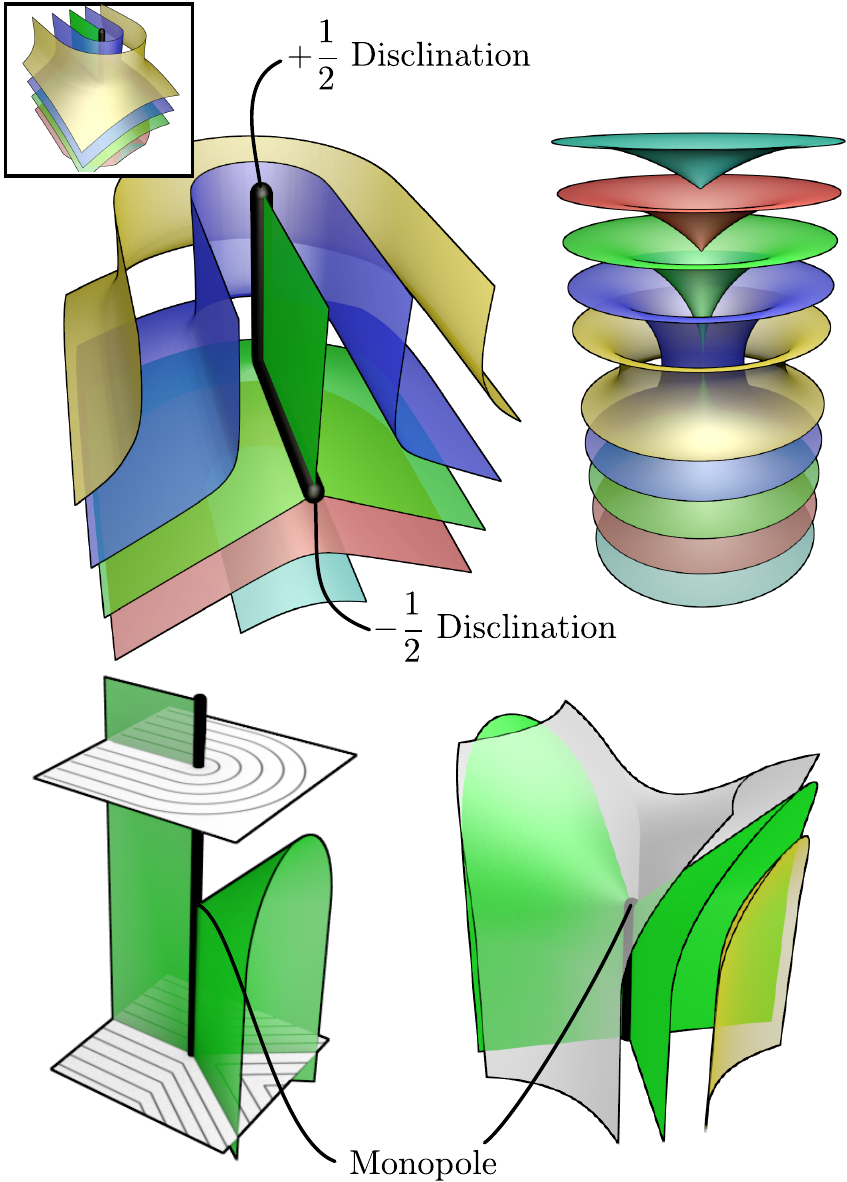}}
	\caption{\label{fig:monopole} 
		(Top Left) A $\ph$ disclination line is converted into a $\mh$ disclination line at a {\sl pincement} (inset -- rear view).
		(Top Right) A toric focal conic domain inserted into an otherwise unperturbed smectic. If we cut this into four by slicing a vertical plane and a perpendicular horizontal plane through the central point defect, the quarter toric domain is exactly the {\sl pincement} illustrated on the left. (Bottom Left) Converting a $\ph$ disclination to a $\mh$ disclination without a bend in the disclination line. The monopole is located where the cone attaches to the disclination. (Bottom Right) Escaping a $-1$ disclination by attaching two half-cones to the layers. 
	}
\end{figure}

This splitting into disclination pairs reveals an essential difference between edge and screw dislocations that is the central issue of this paper. Edge dislocations split into $\ph$/ $\mh$ disclination pairs~\cite{klemanburger,klemanbook},  while screw dislocations break into a pair of $\ph$ disclinations -- how is the topological charge of the disclinations preserved?
In Fig.\ \ref{fig:screw2edge} we illustrate a $b=4$ composite screw bending over to become a composite edge dislocation. Below the transition layers, the bottom layer structure indicates the topology of the composite screw. The transition to the edge dislocation at the top of Fig.\ \ref{fig:screw2edge} preserves this separation of disclinations. However, because the edge dislocation is made of a $\ph$/$\mh$ disclination pair and the screw disclination is made of two $\ph$ disclinations, the transition requires the conversion of a $\ph$ disclination into a $\mh$ disclination.   As we show in Fig.\ \ref{fig:monopole}, it is possible to turn a $\ph$ disclination into a $\mh$. While this geometry reflects the more familiar fact that in a three-dimensional nematic there is only one kind of line defect locally ($\pi_1(\mathbb{R}P^2)=\mathbb{Z}_2$), the existence of smectic layers implies additional structure. 

The layer normal of a surface, $\bf n$, must satisfy the Frobenius integrability condition ${\bf n}\!\cdot\!\left(\nabla\!\times\!{\bf n}\right)=0$ so there can be no twist in a smectic. In addition, smectics must satisfy a geometric `measured' condition that follows from a finite layer thickness. As we show below, these restrictions imply that the transition from $\ph$ to $\mh$ must occur at an identifiable point, a `monopole' sitting on the disclination line. At a generic point $p$ along a smectic disclination one can associate an integer that counts the number of layers, $m_p$ attached to the disclination at that point. This is equal to 1 for a $\ph$ disclination, and 3 for a $\mh$ disclination. Generally, the local winding number of the disclination at $p$ is given by $1-m_p/2$. Because $m_p \geq 0$ by construction, a smectic disclination can have a maximum geometric winding of $+1$, a consequence of Po\'enaru's result concerning the measured condition of smectics~\cite{poenaru}.

Since $m_p$ is an integer valued function along the disclination line, it can only change discretely at specific points. These are the monopoles, and are unique to smectics. No such structure can be defined for a nematic liquid crystals, where a $+q$ profile may be smoothly deformed to a $-q$ profile. As there is no homotopy between windings of $\pm q$ with ${\bf n}$ lying in a plane, such a homotopy must fully explore the groundstate manifold, $\mathbb{R}P^2$, and evolve into the third dimension. While the smectic configuration is a valid nematic texture,
in a nematic such a configuration is not topologically protected from smearing to a smooth transition with twist (consider the standard $\pmh$ transition through a twist disclination). This twist
prohibits the definition of a phase field and consequently the integer-valued invariant $m_p$ is not defined in a nematic.

So how does the $\pmh$ transition occur in a smectic?  In the top row of Fig.~\ref{fig:monopole} we consider such a transition made by cutting open a toric focal conic domain.  In this case the natural $\pi/2$ turn abets the transition from screw to edge but also demonstrates a key feature of the transition: it occurs at a point.  Disclination lines in a smectic are where smectic layers intersect along a line (or end along a line as in the $\ph$ and $+1$ geometries) so, in order to make a transition, a new layer must emerge from the line.  A whole layer adds two leafs and so the geometric winding would change by $-1$ or, conversely, increase by $1$ on removal.  Note that a half-layer (density minimum) joining would amount to two disclination lines joining -- a different beast altogether.  Thus, in order to make the geometric transitions it is necessary to go through a more singular point defect, a critical point of the smectic layers that is also a point defect or monopole.  
This minimalistic layer description in terms of whole sheets and sheets ending on disclination lines is equivalent to a full phase field model, as we show rigorously elsewhere~\cite{toappear}.

More generally, it is not necessary to use a focal conic domain segment to achieve the transition, as illustrated in the bottom images of Fig.~\ref{fig:monopole}. In order to convert a charge-$m$ disclination to a charge-$m'$ one, $2(m-m')$ layers need to be added in the cross-section perpendicular to the disclination line.  Generically, disclination lines do not intersect each other, and so the layers must appear in pairs by forming a three-dimensional half-conical structure. It follows that $m-m'$ is necessarily integer, and nematic order can be maintained. For every integer there are monopole structures carrying that charge. The complete set of rules for the topologically allowed moves is the subject of other work~\cite{toappear}.

With this discussion in mind, we return to the transition from a screw to an edge dislocation, it is fortunate that the transition can be achieved with a $\pi/2$ turn:   an edge dislocation (when the defect line is perpendicular to the displacement) and  a screw dislocation (when the defect line is parallel to the displacement) meet as in Fig.~\ref{fig:screw2edge}.  We see that one disclination in the screw keeps its charge and becomes the companion for the newly created $\mh$. The remaining $\ph$ charge pairs up with a $\ph$ generated through a pinch or {\sl pincement} of the smectic as illustrated in the topmost layer in Fig.\ \ref{fig:screw2edge}.  The topmost full layer merges with the disclination and, at that point the pincement is created.

Experimental observations \cite{williams75} demonstrate the existence of composite screw dislocations.  We determine the energetic favorability of the split-core screw dislocation expanding upon the geometric description by Kl\'eman and coworkers~\cite{meyer10}. The topology of a screw dislocation requires $\phi_{\rm screw}$ at large distances, a solution with vanishing mean curvature $H$. We replace the core with equally-spaced layers (Fig.\ \ref{fig:composite}) built with level sets of $\phi$ specified as the normal evolution of a central helicoid by a distance $\ell$:
${\bf X}_\ell(r,\theta) = \left[r\cos\theta, r\sin\theta, \bbar\,\theta\right] + \ell\, {\bf N}$, where $\bbar\equiv b/(2\pi)$ and ${\bf N}$ is the normal of the central helicoid, ${\bf N}(r,\theta) = \gamma\left[\sin\theta,-\cos\theta,r/\bbar \right]$,
with $\gamma=[1+(r/\bbar)^2]^{\mh}$ normalizing $\bf N$. On the central helicoid $\ell=0$, the mean curvature vanishes identically and the Gaussian curvature is $K(r,\theta) = -\gamma^4/\bbar^2$ so that the two principal curvatures are $\kappa_\pm =\pm \gamma^2/\bbar$.  The largest value of $\vert\kappa\vert = 1/\bbar$ and measures  the inverse distance to the first curvature singularity generated by the normal evolution of the helicoid, $\bbar$. Except for the central helicoid, the core layers are not minimal surfaces, {\sl i.e.} $H\ne 0$, however the compression in the entire core vanishes by construction.

It is amusing to note that this dichotomy of vanishing compression in the core and vanishing curvature in the exterior is reminiscent of the structure of e.g. an Abrikosov flux line~\cite{Abrikosov}: at large distances we have a superconducting phase with vanishing magnetic field, while in the core we have normal metal with penetrating flux.  In the flux line case this is a balance between two linear terms in the London theory.  In our case, the smectic  free energy is nonlinear in $\phi$ but harmonic in the compression strain $u_{zz}=(\nabla\phi)^2-1$ and the mean curvature $H$.  This same unexpected balance between nonlinear strains was first pointed out by Brener and Marchenko in smectic edge dislocations~\cite{BM}.

It is straightforward to calculate the curvature energy of this core structure and we find $F_{\rm core} = 4.66 B \lambda^2$. The core energy is independent of the reduced Burgers scalar $\bbar$ since \eqref{eq:elas_en} reduces to the conformally invariant Willmore energy in the case of equally spaced layers.
This adds to the energy of the exterior region, which has only compression energy since it has zero mean curvature. Substituting the expression for $\phi_{\rm screw}$ into \eqref{eq:elas_en} this energy is found to be $F_{\rm shell}=(\pi/8)B\bbar^2.$

There are two things that we must check at the interface between the core and the shell: 1) whether the layers match at the cylinder of radius $\bbar$, and 2) whether the layers match up smoothly or there is a mismatch between layer normals where the core meets the shell. To fit onto the shell layers, the inner surfaces must intersect the circle of radius $\bbar$ at height $z=0$ at equally-spaced angles, so that the pushoff at distance $\ell$ from the central helicoid intersect the circle at an angle $\pi \ell/(2\bbar)$.  However, it is straightforward but tedious to check that the intersection is at an angle
$\alpha(\tilde\ell)=[\tilde\ell \sqrt{k/(1 + k)}+ \tan^{-1}(\tilde\ell/\sqrt{k+k^2})]
$
where $k^2=1-\tilde \ell ^2$. The difference between $\alpha(\tilde{\ell})$ and $\pi\tilde{\ell}/2$ is greatest at $|\tilde\ell|\approx0.73$, where the two differ by $8\%$.  Thus the core cannot have vanishing compression and also attach continuously to the other leaves and the double helicoidal structure proposed in \cite{achard05} requires this small tweak.
The additional compression energy can be estimated by changing the spacing of the pushoffs in the core so that the distance of the ${\ell}^{\rm th}$ layer from the central helicoid is $\bbar\alpha^{-1}(\pi\tilde{\ell}/2)$ rather than $\ell$. With this adjustment, the core spacing now reads $|\nabla\phi|=\frac{d}{d\tilde{\ell}}[\alpha^{-1}(\pi\tilde{\ell}/2)]$, with  compression energy $\sim .015\,B\bbar^2$, a nonzero but small correction to the compression in the outer region 
quadratic in $\bbar$.  
The bending energy stored in the mismatch between the layer normals carries a delta-function of mean curvature, which will not scale with $\bbar$ due to the conformal invariance of the Willmore energy. Computing the angle deficit $\beta_{\bf N}(\tilde{\ell})=\arccos({\bf N}_c\cdot{\bf N}_s)$, where ${\bf N}_c$ and ${\bf N}_s$ are the core and shell layer normals respectively. Allows us to estimate the total bending energy in the mismatch between layer normals as $[\beta_{\bf N}(\tilde{\ell})]^2$ along the core boundary, giving $\sim 0.79\,B\lambda^2$,  a small correction to the bending energy of the core.  

Even with the modification of the core structure, we find that the energy scales as $F_{\rm composite} \sim B b^2 +C$, as argued in \cite{meyer10}.  For large $b$ this will be smaller than the energy of the traditional, microscopic-core dislocation $F_{\rm standard} \sim Bb^4/\xi^2+\chi \xi^2$, where $\xi$ is the core radius and $\chi$ is the smectic condensation energy density. We can also compare $F_{\rm composite}$ to the energy of a screw dislocation with an elastically melted nematic core~\cite{msk}; minimizing the energy over $\xi$ gives $\xi\propto b$, leading to the scaling $F_{\rm standard}\sim\sqrt{B\chi}b^2$. Deep in the smectic phase, $\chi>B$, and therefore for large enough $b$ the composite screw will have lower energy than the standard melted-core screw. 

We have described the structure of a composite-core dislocations comprised of two disclination lines and their topology.   In particular,  line disclinations in three-dimensional smectics carry a $\mathbb{Z}_2$ topological charge, exactly as in three-dimensional nematics.  Unlike nematics however, ``escape into the third dimension'' is not allowed in smectics and so the homotopy between different winding geometries occurs via higher-order monopoles. The work presented here is valid only for smectic A textures, where the layers have no additional structure,  Smectic C textures require the additional matching of the c-director around the defect which we do not consider. In future work we will complete the classification of defects in smectics by studying the variety of allowable point defects \cite{toappear}.  More generally, the connection between disclinations and dislocations remains an open issue in translationally ordered systems.

\acknowledgments

It is our pleasure to acknowledge penetrating discussions with M. Kl\'{e}man, O.D. Lavrentovich, and J.-F. Sadoc.  This work was supported through NSF Grant DMR1262047 and by a Simons Investigator grant from the Simons Foundation to R.D.K.


\begin{thebibliography}{10}


\bibitem{klemanbook} M. Kl\'eman, \textit{Points, Lines and Walls: In Liquid Crystals, Magnetic Systems and Various Ordered Media}, (John Wiley \& Sons, New York, 1983).

\bibitem{mermin} N.D. Mermin, Rev. Mod. Phys. {\bf 51}, 591-648 (1979).

\bibitem{ack} B.G. Chen, G.P. Alexander, and R.D. Kamien, Proc. Natl. Acad. Sci. {\bf 106}, 15577 (2009).

\bibitem{poenaru} V. Po\'enaru, Commun. Math. Phys. {\bf 80}, 127 (1981).

\bibitem{msk} E.A. Matsumoto, C.D. Santangelo, R.D. Kamien, Interface Focus {\bf 2}  617 (2012).

\bibitem{kl} R.D. Kamien and T.C. Lubensky, Phys. Rev. Lett. {\bf 82} 
2892 (1999).

\bibitem{meyer10} C. Meyer, Y. Nastishin and M. Kl\'{e}man, Phys. Rev. E {\bf 82}, 031704 (2010).

\bibitem{achard05} M.F. Achard, M. Kl\'{e}man, Y. A. Nastishin, and H. T. Nguyen, Eur. Phys. J. E {\bf 16}, 37 (2005).

\bibitem{klbook} M. Kl\'{e}man and O.D. Lavrentovitch, \textit{Soft Matter Physics: An Introduction}, (Springer-Verlag, New York, 2003).

\bibitem{williams75} C. E. Williams, Philos. Mag. {\bf 32}, 313 (1975).

\bibitem{LL} L.D. Landau, E.M. Lifshitz, ,A. M. Kosevich, and L. P. Pitaevski\u{i}  , \textit{Theory of Elasticity (Third Edition)} (Elsevier-Butterworth Heinemann, Oxford, 1986).

\bibitem{London} F. London and H. London, Proc. Roy. Soc. A {\bf 149}, 71 (1935).

\bibitem{dgsmectic} P.-G. de Gennes, Solid State Commun. {\bf 10}, 753-756 (1972).

\bibitem{halseynelson} T.C. Halsey and D.R. Nelson, Phys. Rev. A {\bf 26}, 2840-2853 (1982).

\bibitem{foot} Note that the term ``Burgers vector'' might be more common but, in a smectic, this is only a scalar quantity.  It would be a vector in a crystal with periodicity in more than one direction.
 
\bibitem{BPS} C.D. Santangelo and R.D. Kamien, Phys. Rev. Lett. {\bf 91}, 045506 (2003).

\bibitem{slse} M.Y. Pevnyi, J.V. Selinger, and T.J. Sluckin, Phys. Rev. E {\bf 90}, 032507 (2014).

\bibitem{klemanburger} C.E. Williams and M. Kl\'eman, J. Phys. (Paris) {\bf 36} (C1), C1-315 (1975).

\bibitem{toappear} T. Machon, H. Aharoni, Y. Hu, and R.D. Kamien, {\sl unpublished} (2017).


\bibitem{Abrikosov} A.A. Abrikosov, Zh. Eksp. i Teor. Fiz. {\bf 32}, 1442-1452 (1957); [JETP {\bf 5}, 1174-1182 (1957)].

\bibitem{BM} E.A. Brener and V.I. Marchenko, Phys. Rev. E {\bf 59}, R4752 (1999).



\end{thebibliography}
\end{document}